# Restricted Modulation Freedom Enhances Noise Robustness in Coherent Diffractive Optical Networks

Hyuntae Lim[1] and Kyoungsik Kim[1,*]

[1]*School of Mechanical Engineering, Yonsei University, 50 Yonsei-ro, Seodaemun-gu, Seoul 03722, Republic of Korea.*

[*]Correspondence and requests for materials should be addressed to K.K.
(email: kks@yonsei.ac.kr).

## Abstract

In coherent diffractive optical networks, greater modulation freedom allows more flexible optimization for clean inputs, but its effect on noise robustness and its physical origin remain unclear. We derive an analytical framework that identifies the physical mechanism linking modulation freedom to noise robustness. To establish this connection, we compare a continuous D²NN (C-D²NN) with continuous amplitude and phase modulation and a binary-mask D²NN (BM-D²NN) with binary amplitude modulation. Trained only on clean MNIST data, the seven-layer BM-D²NN has 1.83 percentage points lower clean test accuracy but 32.82 percentage points higher noisy test accuracy under pixel-wise Gaussian noise. This robustness advantage cannot be explained by lower total noise-only intensity at the imaging plane: the BM-D²NN has 3.22 times the noise-only intensity but 8.64 times the clean-signal intensity of the C-D²NN, reducing noise contamination relative to the clean signal. We show that this difference arises from a clean-signal transmission bias generated by spatial correlations between the structured clean-signal field and learned modulation patterns. We quantify this mechanism with a cumulative transmission-bias factor $K$, linking modulation freedom to relative noise contamination and robustness. The C-D²NN exhibits a stronger suppressive bias (more negative $K$), whereas the BM-D²NN exhibits a weaker bias (less negative $\widetilde{K}$), yielding the consistent ordering $K < \widetilde{K}$. Because $K$ requires only clean-data forward passes, it serves as a robustness screening metric before noisy-input simulations.

## Introduction

Coherent wave-optical analog processors—including diffractive deep neural networks ($D^2NNs$) [1], optical reservoir computers [2], and free-space matrix multipliers [3]—perform a range of computational tasks by controlling optical fields through a large number of spatial modulation degrees of freedom. However, the performance of these coherent optical processors can be sensitive to practical perturbations, including fabrication errors, thermal drift, and detector noise [4–8]. In $D^2NN$ hardware, for example, fabrication and experimental implementation errors have been reported to reduce experimentally measured accuracy by more than 10 percentage points relative to numerical predictions [9]. Thus, in addition to achieving high computational performance through greater modulation freedom, it is important to understand how robustly these optical systems operate under practical noise conditions.

Greater modulation freedom broadens the range of realizable optical transformations, allowing the response to clean inputs to be optimized more flexibly [10–12]. However, whether this advantage persists during inference in the presence of stochastic input noise remains unclear. A further question is how restricting modulation freedom affects robustness under noisy inference and what physical mechanism governs the resulting difference. To address these questions, we compare a continuous $D^2NN$ (C-$D^2NN$), which uses continuous amplitude and phase modulation and therefore has greater modulation freedom, with a binary-mask $D^2NN$ (BM-$D^2NN$), in which the modulation amplitude is restricted to 0 (fully opaque) or 1 (fully transmissive), thereby strongly limiting modulation freedom.

After both architectures are trained using only clean data, the BM-$D^2NN$ maintains substantially higher classification accuracy than the C-$D^2NN$ in the presence of Gaussian input noise. For example, for the seven-layer MNIST networks shown in Table I, the BM-$D^2NN$ achieves a noisy-test accuracy of 81.21%, whereas the C-$D^2NN$ achieves 48.39%, corresponding to a 32.82 percentage-point gap. Notably, this robustness advantage does not arise because the BM-$D^2NN$ transmits less noise. In fact, with Gaussian noise alone as input, the total noise-only intensity at the imaging plane in the BM-$D^2NN$ is approximately 3.22 times that in the C-$D^2NN$. By contrast, with clean images alone as input, the total clean-signal intensity at the imaging plane in the BM-$D^2NN$ is approximately 8.64 times that in the C-$D^2NN$. Thus, the difference in robustness between the two architectures is associated not with the absolute intensity of the transmitted noise itself, but with the magnitude of the noise relative to the transmitted clean signal.

We show that this difference originates from how the learned modulation patterns transmit the structured clean-signal field. To quantify this effect, we introduce $K$, a network-level quantity that accumulates the clean-signal transmission bias arising at each modulation layer. In the C-$D^2NN$, greater modulation freedom is associated with a stronger suppressive transmission bias (more negative $K$) for the clean-signal field, whereas the binary-amplitude constraint in the BM-$D^2NN$ results in a substantially weaker bias (less negative $\widetilde{K}$), with $K < \widetilde{K}$ for all 10,000 MNIST test images. Consequently, the BM-$D^2NN$ maintains lower noise contamination relative to the transmitted clean signal and exhibits greater noise robustness. Therefore, $K$ quantifies this transmission-bias mechanism and captures the architecture-dependent difference in noise robustness. Because $K$ can be evaluated from clean-data forward passes alone, it can serve as a robustness screening metric for comparing optical architectures before noisy-input simulations are performed. The BM-D²NN's robustness advantage persists in both zero- and nonzero-mean noise regimes over a broad range of network depths. Both this advantage and the underlying transmission-bias mechanism extend beyond MNIST to Fashion-MNIST.

## I. Clean Accuracy and Noise Robustness under Restricted Modulation Freedom

Both the C-D²NN and BM-D²NN consist of $N$ modulation layers separated by Fresnel free-space propagation and use the same propagation geometry and operator. The key difference between the two architectures is the modulation rule applied to the optical field at each layer. To describe the field evolution common to both architectures, we denote the transmittance of the $n$th modulation layer by $T_n(\mathbf{x})$. The optical field $E_{n-1}^{prop}(\mathbf{x})$ at the $(n-1)$th propagation plane is then modulated by $T_{n-1}(\mathbf{x})$ and Fresnel-propagated to the $n$th propagation plane as

$$E_n^{prop}(\mathbf{x}) = \mathcal{F}\left[\mathcal{H}(\mathbf{k})\mathcal{F}^{-1}\left[T_{n-1}(\mathbf{x})E_{n-1}^{prop}(\mathbf{x})\right]\right], \qquad n \geq 2, \tag{1}$$

where the layer transmittances are

$$T_n^{(C)}(\mathbf{x}) \equiv A_n(\mathbf{x})e^{i2\pi\phi_n(\mathbf{x})}, \qquad T_n^{(BM)}(\mathbf{x}) \equiv B_n(\mathbf{x}) \in \{0,1\}. \tag{2}$$

The Fresnel transfer function is $\mathcal{H}(\mathbf{k}) = e^{ik_0 z}e^{-i\frac{\lambda z}{4\pi}|\mathbf{k}|^2}$ [13], where $k_0 = 2\pi/\lambda$ is the free-space wave number and $\mathbf{k}$ is the transverse wave vector. The two architectures therefore share the same Fresnel propagation operator and geometry but differ in their layer transmittances. The C-D²NN allows continuous control of both the amplitude $A_n(\mathbf{x}) \in [0,1]$ and phase $\phi_n(\mathbf{x})$, providing greater modulation freedom. In contrast, the BM-D²NN restricts $B_n(\mathbf{x})$ at each pixel to 0 or 1, corresponding to fully opaque and fully transmissive pixels, respectively. Thus, our comparison holds the propagation operator and geometry fixed while substantially changing the available modulation freedom. The BM-D²NN architecture is illustrated in Fig. 1. In all simulations, we use a wavelength of $\lambda = 550$nm, a propagation distance of $z = 22$μm, and transverse planes of $M \times M = 200 \times 200$ pixels, where $M$ denotes the number of pixels along one side of the transverse plane. Because binarization is nondifferentiable, we train the BM-D²NN using a straight-through estimator (STE) [14–16]: the forward pass uses mask values binarized at a threshold of 0.5, whereas the backward pass uses a unit gradient through the step function. Both architectures are trained on clean MNIST and Fashion-MNIST data using a mean-squared-error (MSE) loss over 10 class-specific spatial intensity detectors.

Table I compares the clean and noisy test accuracies of the seven-layer C-D²NN and BM-D²NN. On clean MNIST test data, the C-D²NN and BM-D²NN achieve accuracies of 92.75% and 90.92%, respectively, corresponding to a clean-accuracy reduction of only 1.83 percentage points under binary modulation. Under a representative noisy condition generated by adding pixel-wise Gaussian noise $\mathcal{N}(45, 10^2)$ to each clean test image, the C-D²NN accuracy drops to 48.39%, whereas the BM-D²NN retains 81.21%, yielding a 32.82-percentage-point gap between the two architectures.

Thus, constraining the modulation to binary amplitude values rather than continuously varying amplitude and phase produces an empirical trade-off: a modest loss in clean-data accuracy but a substantial gain in robustness to stochastic input noise. To identify the physical origin of this robustness advantage, we separately examine the transmission of the clean signal and the noise-only component to the imaging plane. We first use noise-only inputs to test whether the greater robustness of the BM-D²NN can be explained by reduced absolute noise transmission.

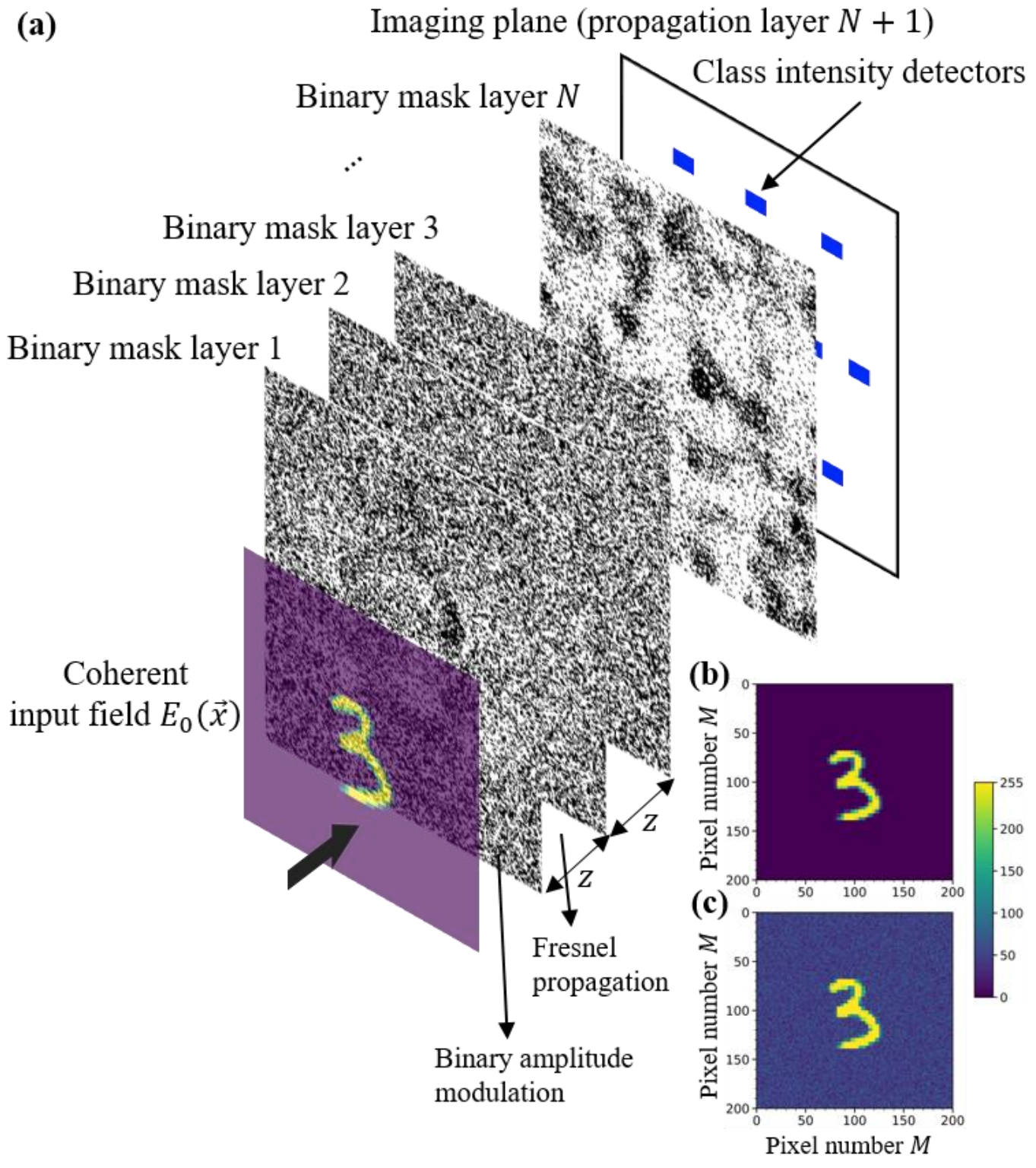

FIG. 1. BM-D²NN architecture and representative clean and noisy inputs. (a) Schematic of the BM-D²NN, in which binary-amplitude masks alternate with Fresnel free-space propagation. (b) Clean MNIST image of the digit 3. (c) Corresponding noisy input obtained by adding pixel-wise Gaussian noise $\mathcal{N}(50, 20^2)$ to the clean input in (b).

TABLE I. Clean and noisy MNIST test accuracies for seven-layer C-D²NN and BM-D²NN models. For the noisy condition, pixel-wise Gaussian noise $\boldsymbol{\mathcal{N}(45, 10^2)}$ was added to each clean MNIST test image.

| **Model** | **MNIST test accuracy (%)** | |
|---|---|---|
| | **Clean** | **Noisy** |
| C-D²NN | 92.75 | 48.39 |
| BM-D²NN | 90.92 | 81.21 |

## II. Noise Robustness despite Greater Noise Transmission

We first test whether the robustness advantage of the BM-D²NN can be explained simply by reduced noise transmission to the imaging plane. If reduced noise transmission were the primary cause of the robustness difference between the two architectures, the BM-D²NN should transmit less noise than the C-D²NN. To test this hypothesis, we use pixel-wise Gaussian noise $\mathcal{N}(m, \sigma^2)$ alone as input, with no clean signal present. For each architecture, we then measure the fraction of the total noise-only intensity at the input that reaches the imaging plane. We define the noise transmission ratio as the total noise-only intensity at the imaging plane divided by the total noise-only intensity at the input. For clean inputs, we analogously define the clean-signal transmission ratio as the total clean-signal intensity at the imaging plane divided by that at the input.

As shown in Fig. 2, under the representative noise-only condition $\mathcal{N}(50, 20^2)$, all filled circles lie above the $y = x$ reference line, showing that the BM-D²NN transmits a larger fraction of the input noise intensity for every model pair. Averaged over the 20 models within each architecture, the mean noise transmission ratio of the BM-D²NN is approximately 3.22 times as large as that of the C-D²NN. Thus, despite its greater noise robustness, the BM-D²NN transmits more, rather than less, noise to the imaging plane, so its robustness advantage cannot be explained by reduced absolute noise transmission.

The clean-signal comparison reveals an even larger architecture-dependent difference. The open circles in Fig. 2 also lie above $y = x$. The mean clean-signal transmission ratio of the BM-D²NN is approximately 8.64 times as large as that of the C-D²NN. For each model pair, the slope of the line from the origin to the corresponding point represents the BM-D²NN transmission ratio relative to that of the C-D²NN. The larger slopes of the clean-signal points therefore provide a direct visual indication that the clean-signal enhancement exceeds the noise-only enhancement. Thus, the BM-D²NN transmits more of both the noise and the clean signal than the C-D²NN, but the increase relative to the C-D²NN is much larger for the clean signal. These results show that noise robustness cannot be understood from absolute noise transmission alone; the magnitude of the noise relative to the transmitted clean signal must also be considered.

To quantify this relative noise contamination, we define the noise contribution as the ratio of the total noise-only intensity to the total clean-signal intensity, both evaluated at the imaging plane:

$$C \equiv \frac{I_{noise,N+1}}{I_{clean,N+1}}, \qquad \tilde{C} \equiv \frac{\tilde{I}_{noise,N+1}}{\tilde{I}_{clean,N+1}} \tag{3}$$

Here, $C$ denotes the noise contribution of the C-D²NN, and throughout the following analysis, a tilde denotes the corresponding BM-D²NN quantity. Accordingly, $\tilde{C}$, $\tilde{I}_{noise,N+1}$, and $\tilde{I}_{clean,N+1}$ denote the BM-D²NN noise contribution, the total noise-only intensity at the imaging plane, and the total clean-signal intensity at the imaging plane, respectively. The subscripts $0$ and $N+1$ denote the input plane and the imaging plane after $N$ modulation layers, respectively. A larger $C$ therefore indicates a larger ratio of noise to transmitted clean-signal intensity at the imaging plane. Because the two architectures are evaluated using the same clean input and the same input-noise distribution, the common input-plane intensities $I_{clean,0}$ and $I_{noise,0}$ are written without tildes. These results show that the robustness difference between the two architectures must be understood in terms of the noise contribution $C$, rather than absolute noise transmission alone. We next analyze how the difference in modulation freedom affects $C$ through the clean-signal transmission bias and its cumulative factor $K$.

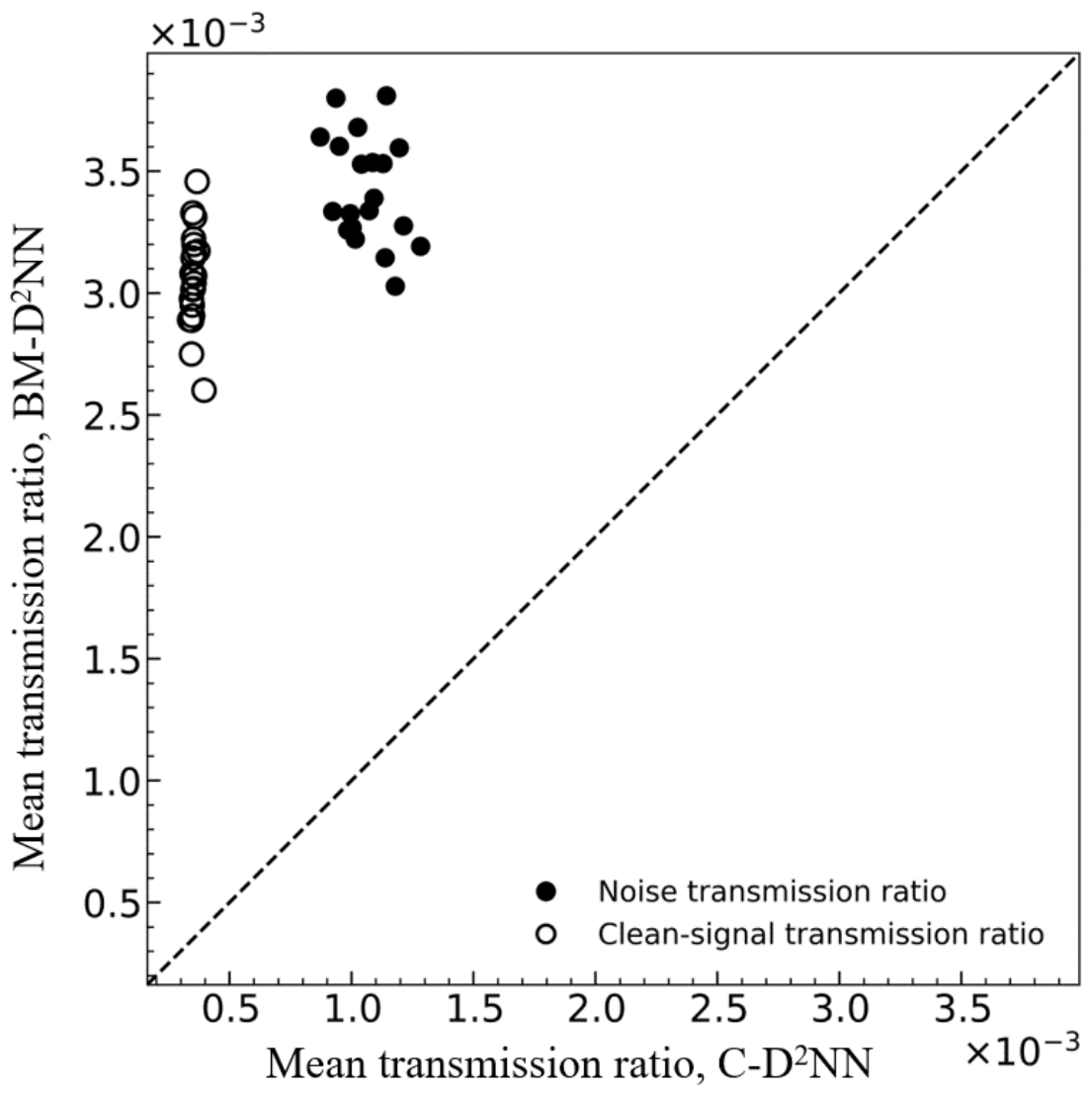

FIG. 2. Paired comparison of noise and clean-signal transmission ratios for 20 independently trained seven-layer C-D²NN/BM-D²NN model pairs. Filled circles denote noise transmission ratios averaged over 10,000 independent noise fields whose pixel values are drawn from $\mathcal{N}(50, 20^2)$; open circles denote clean-signal transmission ratios averaged over all 10,000 clean MNIST test images. The dashed line indicates equal transmission ratios for the two architectures.

## III. Clean-Signal Transmission Bias Links Modulation Freedom to Noise Contribution

To identify the physical origin of the noise contribution $C$, we examine why the clean signal and the noise-only component are transmitted differently through the trained networks. The central distinction is that the structured clean-signal field generated by a clean image can become spatially correlated with the learned modulation patterns formed through clean-data training. The noise-only field lacks this systematic spatial correlation.

To quantify this difference, we first normalize the clean-signal intensity profile incident on the $n$th modulation layer by its spatial mean:

$$\rho_{clean,n}(\mathbf{x}) = \frac{\left|E_{clean,n}^{prop}(\mathbf{x})\right|^2}{\langle\left|E_{clean,n}^{prop}(\mathbf{x})\right|^2\rangle_{\mathbf{x}}}, \qquad \tilde{\rho}_{clean,n}(\mathbf{x}) = \frac{\left|\tilde{E}_{clean,n}^{prop}(\mathbf{x})\right|^2}{\langle\left|\tilde{E}_{clean,n}^{prop}(\mathbf{x})\right|^2\rangle_{\mathbf{x}}} \tag{4}$$

By construction,

$$\langle \rho_{clean,n}(\mathbf{x}) \rangle_{\mathbf{x}} = \langle \tilde{\rho}_{clean,n}(\mathbf{x}) \rangle_{\mathbf{x}} = 1. \tag{5}$$

This normalization removes the overall intensity scale of the clean signal and retains only its spatial distribution. Thus, $\rho_{clean,n}(\mathbf{x}) > 1$ identifies a position where the local clean-signal intensity exceeds its spatial mean at that layer.

We next define a layerwise transmission-bias factor to quantify how the spatial distribution of the clean signal interacts with the learned transmission pattern. For the C-D²NN, the local intensity transmittance is $A_n^2(\mathbf{x})$, giving

$$\kappa_n \equiv \frac{Cov\left(A_n(\mathbf{x})^2, \rho_{clean,n}(\mathbf{x})\right)}{\left(\mu_{n,A}\right)^2 + \left(\sigma_{n,A}\right)^2} = \frac{\langle (A_n)^2 \rho_{clean,n} \rangle_{\mathbf{x}}}{\langle (A_n)^2 \rangle_{\mathbf{x}}} - 1. \tag{6}$$

For the BM-D²NN, $B_n^2 = B_n$, and therefore

$$\tilde{\kappa}_n \equiv \frac{Cov\left(B_n(\mathbf{x}), \tilde{\rho}_{clean,n}(\mathbf{x})\right)}{\tilde{\mu}_{n,B}} = \frac{\langle B_n \tilde{\rho}_{clean,n} \rangle_{\mathbf{x}}}{\langle B_n \rangle_{\mathbf{x}}} - 1. \tag{7}$$

Here, $\mu_{n,A} = \langle A_n \rangle_{\mathbf{x}}$ and $\sigma_{n,A}$ are the spatial mean and standard deviation of $A_n$, respectively, while $\tilde{\mu}_{n,B} = \langle B_n \rangle_{\mathbf{x}}$ is the spatial mean of the binary mask $B_n$. The factor $\kappa_n$ quantifies the effect of preferential spatial overlap between the clean-signal intensity profile arriving at the $n$th modulation layer and the learned transmission pattern at that layer. Two transmission patterns with the same mean transmission can transmit different amounts of the clean signal depending on whether regions of high clean-signal intensity preferentially coincide with high- or low-transmission regions.

A spatially uniform clean-signal intensity profile provides a simple reference case. In this limit, $\rho_{clean,n}(\mathbf{x}) = 1$ or $\tilde{\rho}_{clean,n}(\mathbf{x}) = 1$ everywhere, so no preferential spatial overlap can arise between the clean-signal intensity profile and the learned transmission pattern. The transmitted clean-signal intensity is then determined solely by the mean transmission of the layer, giving $\kappa_n = 0$ or $\tilde{\kappa}_n = 0$. Thus, a zero transmission-bias factor corresponds to the absence of any additional spatial bias beyond the mean-transmission baseline. Conversely, when the clean-signal intensity profile is spatially structured, preferential overlap with the learned transmission pattern can produce a nonzero transmission bias.

The sign of $\kappa_n$ specifies the direction of this bias. When $\kappa_n < 0$, regions of high clean-signal intensity preferentially coincide with lower-transmission regions. Equivalently, bright regions of the clean-signal field are spatially anti-aligned with high-transmission regions of the learned pattern. The transmitted clean signal is therefore smaller than that expected from the mean transmission alone, so a negative $\kappa_n$ represents a suppressive transmission bias. Conversely, when $\kappa_n > 0$, regions of high clean-signal intensity preferentially coincide with high-transmission regions, increasing clean-signal transmission above the mean-transmission baseline. The magnitude $|\kappa_n|$ quantifies the strength of the spatial bias, while its sign determines whether the bias suppresses or enhances clean-signal transmission. The same interpretation applies to $\tilde{\kappa}_n$.

This layerwise bias is directly connected to modulation freedom. The C-D²NN can control continuously varying amplitude and phase and therefore has greater freedom to spatially shape the structured clean-signal field. For the trained networks, this greater modulation freedom is associated with stronger spatial anti-alignment between the clean-signal intensity distribution and the learned transmission pattern, resulting in more negative transmission-bias factors. By contrast, the binary-amplitude constraint in the BM-D²NN limits this spatial shaping freedom and produces a weaker suppressive bias. Thus, the C-D²NN exhibits a stronger suppressive clean-signal transmission bias, reflected in generally smaller values of $\kappa_n$, whereas the BM-D²NN exhibits a weaker bias, reflected in larger values of $\tilde{\kappa}_n$. Accordingly, the layerwise ordering $\kappa_n < \tilde{\kappa}_n$ holds with high probability in the trained networks.

Because these layerwise biases act sequentially through the network, they accumulate multiplicatively. We therefore define the cumulative transmission-bias factors over $N$ modulation layers as

$$K = \prod_{n=1}^{N}(1+\kappa_n) - 1, \qquad \tilde{K} = \prod_{n=1}^{N}(1+\tilde{\kappa}_n) - 1 \tag{8}$$

The factor $1 + K$ represents the multiplicative change in clean-signal transmission through the entire network relative to the value expected from the layerwise mean transmissions alone. Accordingly, $K = 0$ indicates no cumulative transmission bias, whereas $K < 0$ indicates cumulative suppression of the clean signal by preferential spatial overlap.

Applying this decomposition to the clean-signal intensity gives

$$I_{clean,N+1} = I_{clean,0}(1+K)\prod_{n=1}^{N}\left[\left(\mu_{n,A}\right)^2 + \left(\sigma_{n,A}\right)^2\right] \tag{9}$$

for the C-D²NN and

$$\tilde{I}_{clean,N+1} = I_{clean,0}\left(1+\tilde{K}\right)\prod_{n=1}^{N}\tilde{\mu}_{n,B} \tag{10}$$

for the BM-D²NN. Here, $I_{clean,N+1}$ and $\tilde{I}_{clean,N+1}$ denote the total clean-signal intensities at the imaging plane for the C-D²NN and BM-D²NN, respectively. These expressions separate clean-signal transmission into two physically distinct contributions: a mean-transmission contribution, given by the corresponding product of layerwise mean-transmission factors, and a cumulative transmission-bias contribution, given by $1+K$ or $1+\tilde{K}$. Thus, clean-signal transmission depends not only on the mean transmission of each layer but also on the spatial relation between the learned transmission pattern and the structured clean-signal field. The derivation of Eqs. (9) and (10) is provided in Supplemental Material S1.10.

The noise-only field behaves differently because it lacks this spatial correlation with the learned modulation patterns. Under the phase-mixing conditions of our system, its propagation can be characterized using the statistical moments of the propagated field. In our system, $z/\lambda = 40 \gg 1$ provides substantial phase mixing during Fresnel propagation, while the $M \times M = 200 \times 200$ transverse grid provides a large spatial ensemble with $M^2 \gg 1$. Under these conditions, the spatial distribution of the noise-only complex field values after each Fresnel propagation is well approximated by a proper complex Gaussian (PCG) distribution [17,18]. The approximately circularly symmetric distributions in Fig. 3 support the properness assumption underlying the PCG description. Together with the large-pixel-count limit, the PCG approximation reduces noise-only propagation to macroscopic statistical moments, enabling a compact analytical expression for the BM-D²NN and a corresponding lower-bound form for the C-D²NN.

For the BM-D²NN, the total noise-only intensity at the imaging plane is

$$\tilde{I}_{noise,N+1} \approx I_{noise,0} \prod_{n=1}^{N} \tilde{\mu}_{n,B} \,. \tag{11}$$

Because $B_n^2 = B_n$, the noise-only transmission is governed by the product of the layerwise binary fill fractions. For the C-D²NN, however, the phase-dependent complex mean term prevents an exact closed-form expression for the total noise-only intensity based solely on macroscopic amplitude statistics. Retaining the contribution directly determined by these statistics gives the lower-bound form

$$I_{noise,N+1} \geq I_{noise,0} \prod_{n=1}^{N} \left[ \left(\mu_{n,A}\right)^2 + \left(\sigma_{n,A}\right)^2 \right]. \tag{12}$$

The structural distinction between the clean signal and the noise-only component is now explicit. The noise-only expressions above contain the layerwise mean-transmission factors, whereas the clean-signal expressions contain the corresponding factors together with the additional cumulative factor $1 + K$ or $1 + \widetilde{K}$, which accounts for the spatial bias produced by the structured clean-signal field. Derivations of Eqs. (11) and (12) and details of the PCG approximation are provided in Supplemental Material S1–S3.

Substituting the clean-signal and noise-only intensity expressions into the definition of the noise contribution gives

$$C \geq \frac{1}{1+K} \frac{I_{noise,0}}{I_{clean,0}}, \qquad \tilde{C} = \frac{1}{1+\widetilde{K}} \frac{I_{noise,0}}{I_{clean,0}}. \tag{13}$$

Since the input ratio $I_{noise,0}/I_{clean,0}$ is common to both architectures, the architecture-dependent ordering of the noise contribution is governed by the cumulative transmission-bias factors, giving the central prediction

$$K < \widetilde{K} \Rightarrow \tilde{C} < C. \tag{14}$$

This relation connects modulation freedom to noise robustness. The continuous amplitude and phase of the C-D²NN provide greater spatial shaping freedom, and in the trained networks this freedom produces a stronger suppressive spatial bias, making $K$ more negative. The binary-amplitude constraint in the BM-D²NN weakens this suppressive bias, leaving $\widetilde{K}$ less negative. Consequently, the BM-D²NN preserves the clean signal more effectively relative to the noise-only component and therefore exhibits a smaller noise contribution, $\tilde{C}$. Its robustness advantage thus arises not from reduced absolute noise transmission, but from the relative preservation of the clean signal under restricted modulation freedom.

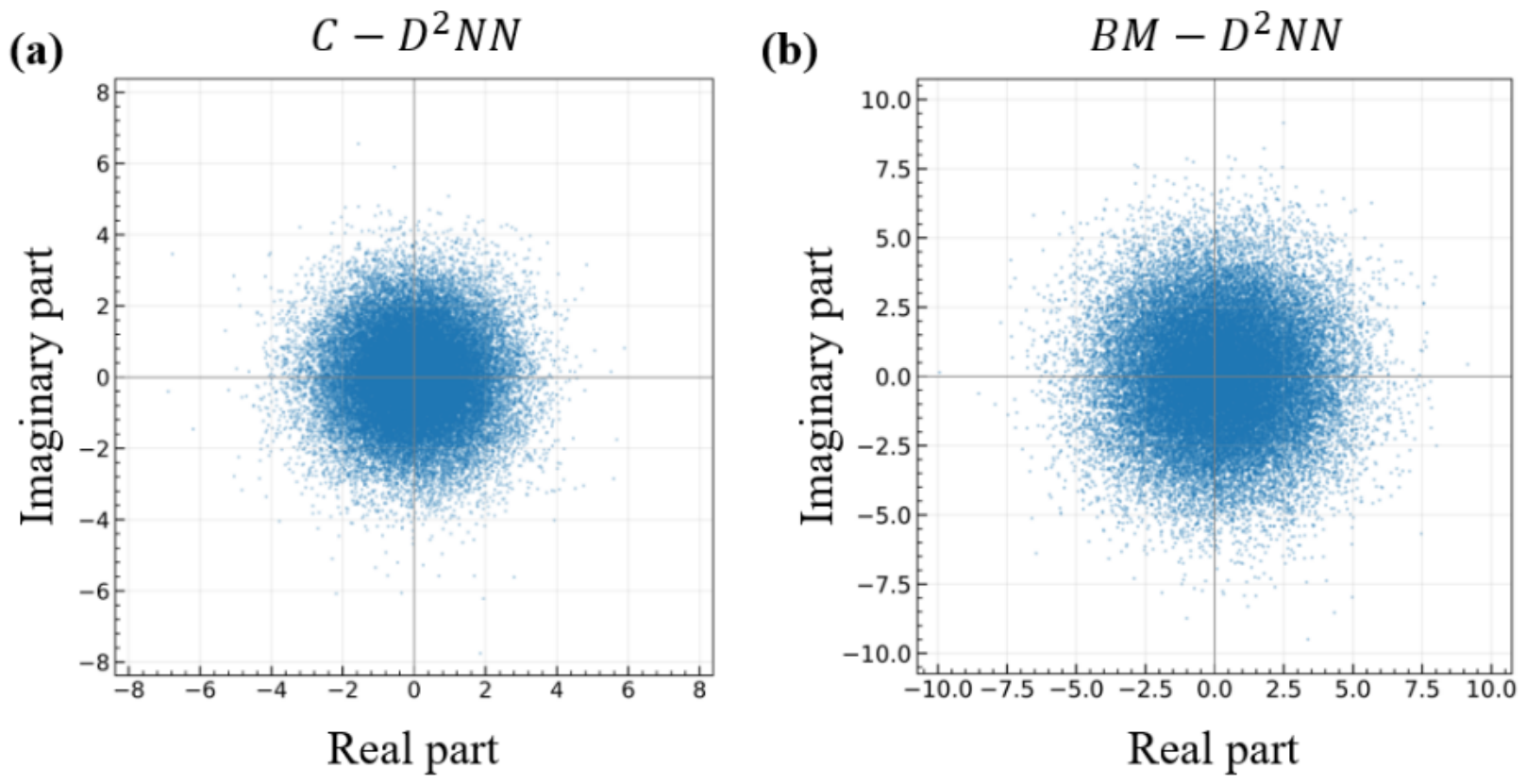

FIG. 3. Real–imaginary scatter plots of the imaging-plane complex field values produced by a noise-only input for (a) the C-D²NN and (b) the BM-D²NN under pixel-wise Gaussian noise $\mathcal{N}(50, 20^2)$.

## IV. Numerical Validation of the Transmission-Bias Mechanism

To numerically validate the transmission-bias mechanism derived in the preceding section, we calculated the layerwise transmission-bias factors $\kappa_n$ and $\tilde{\kappa}_n$, together with the cumulative factors $K$ and $\widetilde{K}$, for 10,000 clean MNIST test images. This analysis tests whether the transmission-bias ordering observed at individual layers accumulates consistently into the predicted full-network ordering $K < \widetilde{K}$.

As shown in Fig. 4(a), $\kappa_n < \tilde{\kappa}_n$ holds for more than 90% of the test images, and in the early layers this ordering holds for nearly all samples. In particular, before reaching the first modulation layer, both architectures undergo the same Fresnel propagation; hence, for the same clean input, $\rho_{clean,1}(\mathbf{x}) = \tilde{\rho}_{clean,1}(\mathbf{x})$. The first-layer ordering $\kappa_1 < \tilde{\kappa}_1$ therefore directly reflects how the two learned modulation patterns differ in their preferential overlap with the same incident clean-signal profile. Numerically, the mean first-layer factors are $\kappa_1 = -0.4776$ and $\tilde{\kappa}_1 = -0.1042$, with the largest difference between the two architectures occurring in the early layers.

These layerwise differences accumulate across the network. In Fig. 4(b), the distributions of $K$ and $\widetilde{K}$ are clearly separated, with $K < \widetilde{K}$ for all 10,000 MNIST test images. The mean cumulative factors, $K = -0.635$ and $\widetilde{K} = -0.132$, show that the C-D²NN develops a substantially stronger suppressive cumulative bias. Thus, the ordering $K < \widetilde{K}$ arises from the accumulation of a consistent signed transmission-bias ordering across multiple layers rather than from a single-layer anomaly, indicating that the larger clean-signal transmission of the BM-D²NN is a network-wide structural effect.

Next, we test whether this intensity-level mechanism is reflected in actual noisy classification performance across independently trained models with different initializations. As shown in Fig. 4(c), the BM-D²NN models cluster in a region of lower noise contribution and lower error rate, whereas the C-D²NN models occupy a region where both quantities are higher. A global regression pooling both architectures yields $R^2 = 0.964$, while $C$ completely separates the two architecture families (AUC = 1.00). These results show that the noise contribution captures the ordering of noisy classification performance across architectures. By contrast, the correlations within each architecture are weak, with $R^2 = 0.007$ for the C-D²NN and $R^2 = 0.21$ for the BM-D²NN. Thus, $C$ should be interpreted not as a universal predictor of the exact error rate of individual models, but as a cross-architecture quantity for comparing the relative noise robustness of distinct architecture families.

Because the preceding section links $C$ to $K$ through the factor $1/(1+K)$, the cumulative transmission-bias factor can also be used for architecture-level robustness comparisons. Specifically, $K$ and $\widetilde{K}$ can be evaluated as

$$K = \frac{I_{clean,N+1}}{I_{clean,0}\prod_{n=1}^{N}\left[\left(\mu_{n,A}\right)^2 + \left(\sigma_{n,A}\right)^2\right]} - 1 \tag{15}$$

$$\widetilde{K} = \frac{\tilde{I}_{clean,N+1}}{I_{clean,0}\prod_{n=1}^{N}\left[\tilde{\mu}_{n,B}\right]} - 1. \tag{16}$$

These expressions require only quantities obtainable from clean-data forward passes: the total clean-signal intensity at the input, the total clean-signal intensity at the imaging plane, and the layerwise amplitude statistics. Accordingly, $K$ can serve as a robustness screening metric for comparing the expected robustness ordering of candidate architectures before noisy-input simulations are performed.

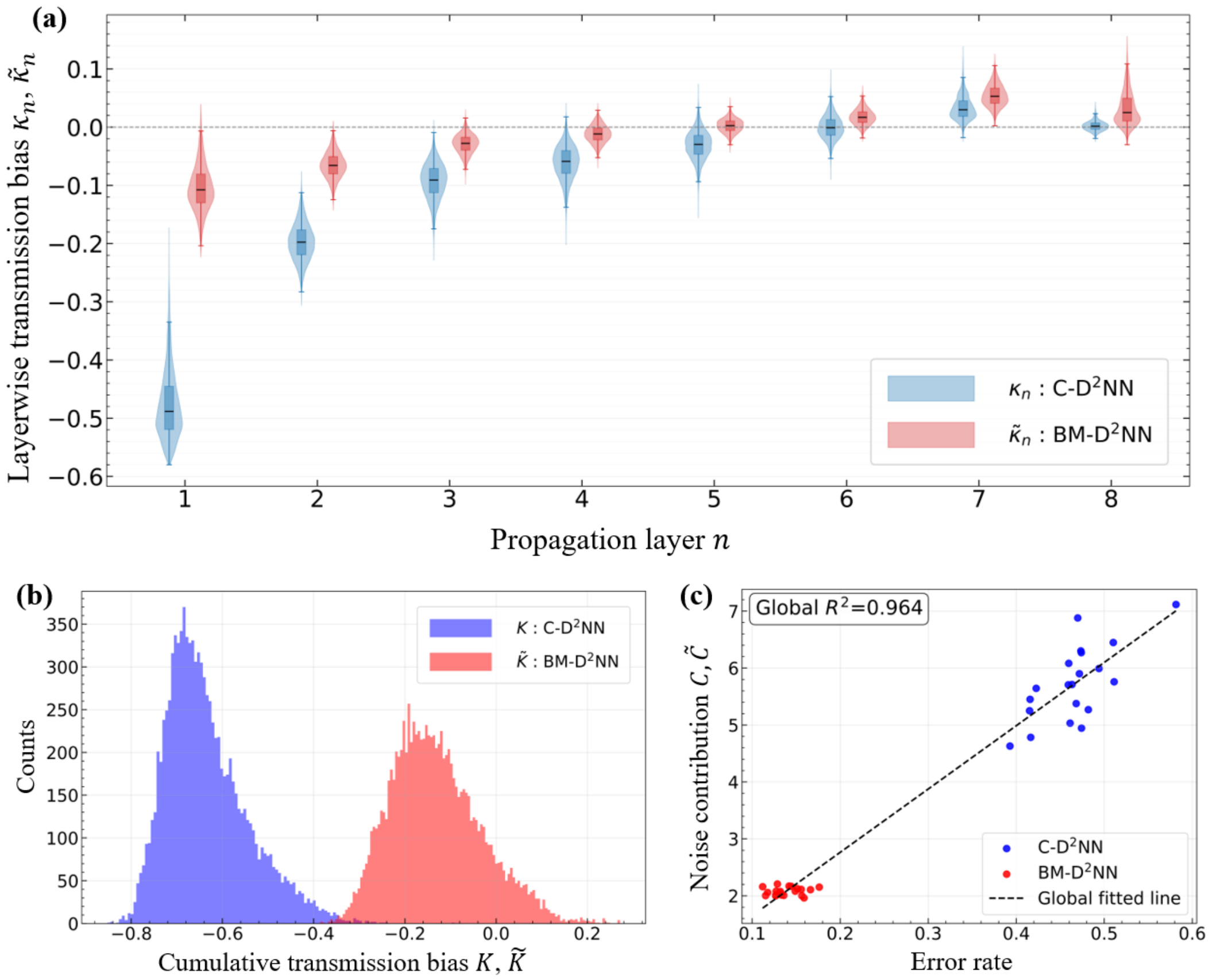

FIG. 4. Numerical validation of the transmission-bias mechanism. (a) Layerwise distributions of the transmission-bias factors $\kappa_n$ and $\tilde{\kappa}_n$ over 10,000 clean MNIST test images. For the seven-layer networks, $n = 8$ denotes the imaging plane after the final Fresnel propagation. (b) Histograms of the corresponding cumulative transmission-bias factors $K$ and $\tilde{K}$ for the same test images. (c) Noise contribution versus noisy-test error rate for 20 independently trained seven-layer models of each architecture under pixel-wise Gaussian noise $\mathcal{N}(50, 20^2)$; the dashed line is a global linear fit pooling both architectures.

## V. Generality across Noise Conditions, Network Depths, and Datasets

To assess the generality of this robustness advantage, we examine whether it extends beyond the representative noise conditions and the seven-layer architecture considered above. Both architectures are trained using only clean data and are evaluated on MNIST test images corrupted by additive pixel-wise Gaussian noise $\mathcal{N}(m, \sigma^2)$. The nonzero-mean regime ($m > 0$) represents additive Gaussian perturbations with a positive offset that can arise from dark-current and electronic offset processes, while thermal and readout processes contribute additional fluctuations [19–22]. The zero-mean regime ($m = 0$) instead represents unbiased Gaussian fluctuations and can be viewed as a simplified Gaussian representation of centered shot-noise fluctuations at high photon flux [19,21–23]. Accordingly, the $(m, \sigma)$ sweep examines whether the robustness advantage extends across physically distinct noise regimes rather than merely across different noise magnitudes.

For each $(m, \sigma)$ condition in Figs. 5(a) and 5(b), the two architectures are evaluated under the same Gaussian noise distribution, so the total noise-only intensity at the input, $I_{noise,0} = M^2(m^2 + \sigma^2)$, is common to both architectures. Because $K$ and $\widetilde{K}$ are determined entirely from clean-signal propagation, they are independent of $m$ and $\sigma$ for fixed trained networks and clean inputs. Therefore, the common input ratio $I_{noise,0}/I_{clean,0}$ varies with $m$ and $\sigma$, whereas the ordering $K < \widetilde{K}$ remains unchanged. Equation (14) then predicts $\tilde{C} < C$ throughout the tested $(m, \sigma)$ range, consistent with the higher noisy-test accuracy of the BM-D²NN. The advantage also persists for zero-mean noise, showing that it is not confined to nonzero-mean noise. At fixed $m = 50$ in Figs. 5(e) and 5(f), the $m^2$ contribution remains larger than the $\sigma^2$ contribution throughout the investigated range. The resulting weaker dependence of $I_{noise,0}$ on $\sigma$ is consistent with the horizontal banding of the noisy-test accuracy maps in Figs. 5(e) and 5(f).

The qualitative robustness ordering is also preserved as the network depth changes. Across the depth-dependent results in Figs. 5(c)–5(f), the BM-D²NN maintains higher noise robustness over the investigated range of $N$. Because $K$ and $\widetilde{K}$ accumulate layerwise transmission biases across the network, their numerical values can change with $N$. Nevertheless, the robustness advantage is maintained over a broad range of network depths, showing that the observed behavior is not specific to the seven-layer configuration.

Finally, to test whether this mechanism generalizes beyond MNIST, we extend the analysis to Fashion-MNIST. The BM-D²NN maintains higher noisy-test accuracy across the tested $m$, $\sigma$, and $N$ ranges. Across 20 pairs of seven-layer models independently trained on Fashion-MNIST, the mean clean-signal transmission ratio of the BM-D²NN is 9.63 times that of the C-D²NN, whereas its mean noise-only transmission ratio is only 1.90 times as large. Consistently, for one representative model pair, the mean values of the cumulative factors are $K = -0.871$ and $\widetilde{K} = -0.406$, with $K < \widetilde{K}$ for all 10,000 Fashion-MNIST test images (Supplemental Material S6.1–S6.3).

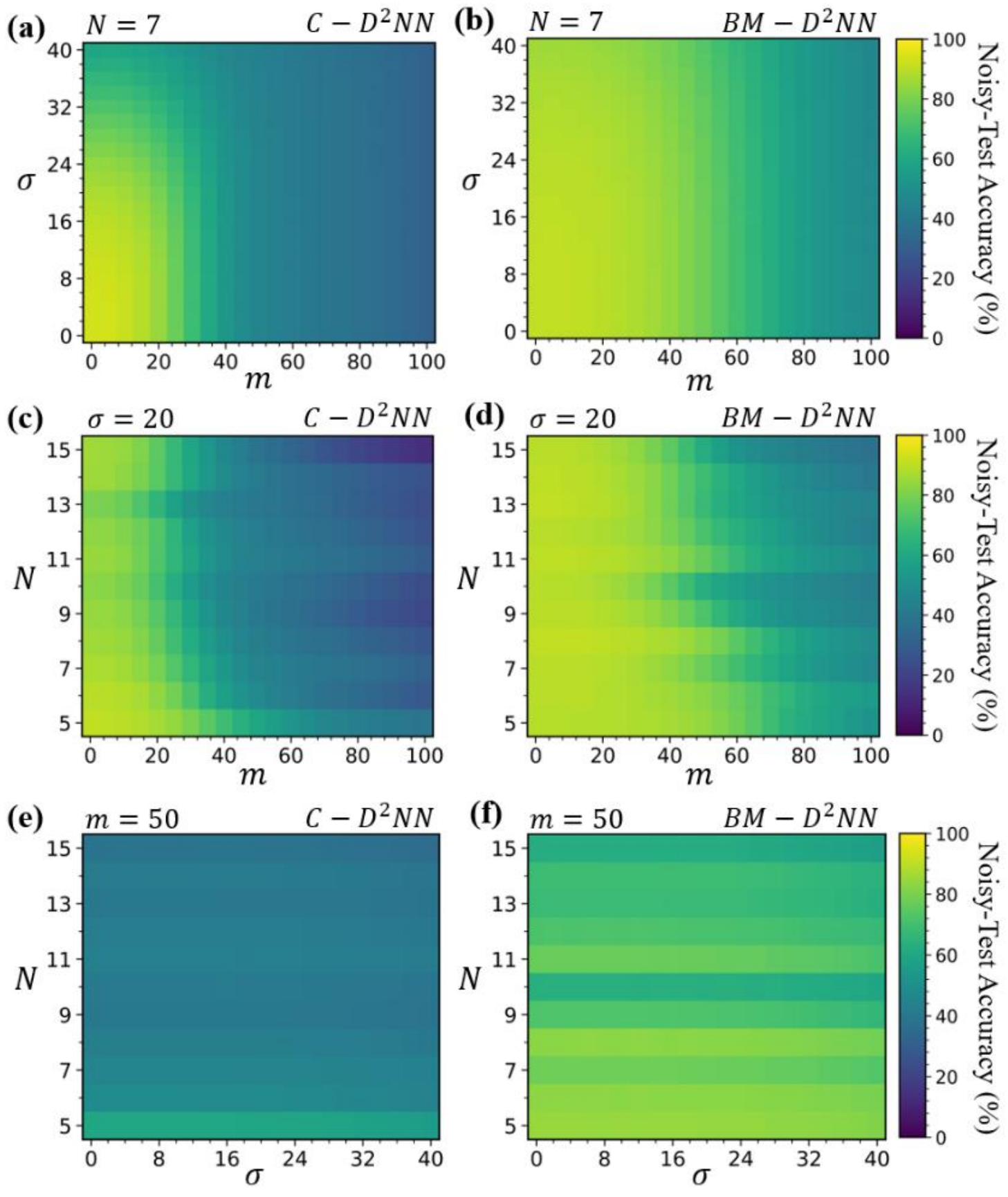

FIG. 5. Noisy MNIST test accuracy under additive pixel-wise Gaussian noise $\mathcal{N}(m, \sigma^2)$ across noise parameters and network depth. The sampled ranges are $m \in [0, 100]$, $\sigma \in [0, 40]$, and $N \in [5, 15]$, where $N$ denotes the number of modulation layers. (a,b) Accuracy as a function of $m$ and $\sigma$ at fixed $N = 7$. (c,d) Accuracy as a function of $m$ and $N$ at fixed $\sigma = 20$. (e,f) Accuracy as a function of $\sigma$ and $N$ at fixed $m = 50$.

## Discussion

The central physical implication of our results is that noise robustness depends on how effectively the clean signal is preserved relative to the noise-only component, rather than on noise transmission alone. Learned modulation patterns can develop spatial correlations with the structured clean-signal field, whereas the noise-only field lacks this systematic spatial correlation. The modulation rule can therefore affect the transmission of the clean signal and the noise-only component differently. In the C-D²NN and BM-D²NN, restricting the modulation to binary amplitude reduces the suppressive clean-signal transmission bias and thereby preserves the clean signal more effectively relative to the noise-only component. Quantitatively, the BM-D²NN exhibits an 8.64-fold enhancement in clean-signal transmission, compared with a 3.22-fold enhancement in noise-only transmission. The greater robustness of the BM-D²NN is therefore best understood as a consequence of this relative clean-signal preservation.

The higher transmission ratios of both the noise-only component and the clean signal in the BM-D²NN can be understood from the cumulative products of the layerwise mean-transmission factors in Eqs. (9)–(12). Within each architecture, the same cumulative product appears in both the clean-signal and noise-only expressions, providing a common baseline for the two ratios. Explicitly, the cumulative products are $\prod_n \langle A_n^2 \rangle_{\mathbf{x}} = \prod_n \left[ \left(\mu_{n,A}\right)^2 + \left(\sigma_{n,A}\right)^2 \right]$ for the C-D²NN and $\prod_n \langle B_n^2 \rangle_{\mathbf{x}} = \prod_n \tilde{\mu}_{n,B}$ for the BM-D²NN. For all 20 independently trained seven-layer model pairs, these products satisfy $\prod_n \langle A_n^2 \rangle_{\mathbf{x}} < \prod_n \langle B_n^2 \rangle_{\mathbf{x}}$, and the ensemble mean of the BM-D²NN product is 3.34 times that of the C-D²NN product. This 3.34-fold ratio is close to the 3.22-fold noise-transmission enhancement observed in Fig. 2, supporting the interpretation that the cumulative mean-transmission factors provide the dominant baseline for this enhancement. This squared-product ordering does not arise simply because the BM-D²NN has larger modulation amplitudes. By contrast, the corresponding products of the unsquared layerwise mean amplitudes satisfy $\prod_n \langle B_n \rangle_{\mathbf{x}} < \prod_n \langle A_n \rangle_{\mathbf{x}}$ for all 20 model pairs. This reversal can be understood from how squaring affects the local intensity transmittance: intermediate continuous amplitudes satisfy $A_n^2 < A_n$ for $0 < A_n < 1$, whereas binary amplitudes satisfy $B_n^2 = B_n$ and are therefore unchanged by squaring. Thus, because optical intensity depends on the squared field amplitude, the different responses of continuous and binary amplitudes to squaring accounts for the observed difference in mean-transmission factors. The resulting higher mean-transmission baseline in the BM-D²NN underlies its greater optical throughput for both the noise-only component and the clean signal. For the clean signal, Eqs. (9) and (10) contain the additional cumulative transmission-bias contribution. Because $K < \widetilde{K}$, the weaker suppressive transmission bias in the BM-D²NN further increases its clean-signal transmission advantage over the C-D²NN, causing the clean-signal enhancement to exceed the corresponding noise-only enhancement.

The transmission-bias mechanism also makes the cumulative transmission-bias factor $K$ useful as an architecture-level robustness screening metric. Because $K$ and $\widetilde{K}$ can be evaluated using only clean-data forward passes, they provide a practical way to compare the expected robustness of candidate architectures without first performing noisy-input simulations. More generally, each layerwise transmission-bias factor is defined through the spatial correlation between the structured clean-signal intensity profile and the layer's intensity transmittance. Its definition is therefore not tied to the binary-amplitude or continuous amplitude-and-phase modulation rules considered here. The same construction can be applied to other modulation rules, including multilevel-amplitude and quantized amplitude-and-phase modulation. A complementary extension is to examine other noise models, including Poisson-distributed and spatially correlated noise, to determine whether the $K$-based robustness ordering continues to hold beyond the Gaussian input-noise regime.

The higher clean-signal transmission of the BM-D²NN also has a hardware implication distinct from this robustness mechanism. When detector noise is a limiting factor, a larger optical signal at the imaging plane can improve detector SNR or reduce the integration time needed to achieve a target SNR [10,20,24,25]. More broadly, our results suggest that modulation freedom should not be treated simply as a resource to maximize, but as an architecture-level design variable that should be chosen to balance clean-data performance and robustness. This transmission-bias perspective may also extend to other coherent optical processors in which structured coherent fields interact with learned spatial modulation.

## Conclusion

We have derived an analytical framework that identifies a physical mechanism linking restricted modulation freedom to noise robustness in coherent diffractive optical networks. Compared with the C-D$^2$NN, the BM-D$^2$NN develops a weaker suppressive clean-signal transmission bias under the binary-amplitude constraint. Consequently, its cumulative factor $\widetilde{K}$ is less negative, and the clean signal is better preserved relative to the noise-only component. This mechanism explains why the BM-D$^2$NN exhibits a lower noise contribution and greater noise robustness despite transmitting a greater total noise-only intensity. The robustness advantage persists across broad ranges of Gaussian noise conditions and network depths. The underlying transmission-bias mechanism extends beyond MNIST to Fashion-MNIST, where the clean-signal transmission enhancement of the BM-D$^2$NN relative to the C-D$^2$NN exceeds the corresponding noise-only transmission enhancement, and the cumulative ordering $K < \widetilde{K}$ is reproduced. Evaluating $K$ using clean data alone also makes the transmission-bias framework useful for architecture-level robustness screening. Thus, the clean-signal transmission-bias framework links modulation freedom to noise robustness and provides a practical basis for balancing clean-data performance and noise robustness in the design of coherent optical architectures.

## References

[1] X. Lin, Y. Rivenson, N. T. Yardimci, M. Veli, Y. Luo, M. Jarrahi, and A. Ozcan, All-optical machine learning using diffractive deep neural networks, Science **361**, 1004 (2018).

[2] M. Rafayelyan, J. Dong, Y. Tan, F. Krzakala, and S. Gigan, Large-scale optical reservoir computing for spatiotemporal chaotic systems prediction, Phys. Rev. X **10**, 041037 (2020).

[3] J. Spall, X. Guo, T. D. Barrett, and A. I. Lvovsky, Fully reconfigurable coherent optical vector–matrix multiplication, Opt. Lett. **45**, 5752 (2020).

[4] M. Y.-S. Fang, S. Manipatruni, C. Wierzynski, A. Khosrowshahi, and M. R. DeWeese, Design of optical neural networks with component imprecisions, Opt. Express **27**, 14009 (2019).

[5] G. L. Zhang, B. Li, Y. Zhu, T. Wang, Y. Shi, X. Yin, C. Zhuo, H. Gu, T.-Y. Ho, and U. Schlichtmann, Robustness of neuromorphic computing with RRAM-based crossbars and optical neural networks, in *Proceedings of the 26th Asia and South Pacific Design Automation Conference (ASP-DAC '21)* (Association for Computing Machinery, New York, 2021), pp. 853–858.

[6] H. Zhang, J. Thompson, M. Gu, X. D. Jiang, H. Cai, P. Y. Liu, Y. Shi, Y. Zhang, M. F. Karim, G. Q. Lo, X. Luo, B. Dong, L. C. Kwek, and A. Q. Liu, Efficient on-chip training of optical neural networks using genetic algorithm, ACS Photonics **8**, 1662 (2021).

[7] G. Mourgias-Alexandris, M. Moralis-Pegios, A. Tsakyridis, S. Simos, G. Dabos, A. Totovic, N. Passalis, M. Kirtas, T. Rutirawut, F. Y. Gardes, A. Tefas, and N. Pleros, Noise-resilient and high-speed deep learning with coherent silicon photonics, Nat. Commun. **13**, 5572 (2022).

[8] Z. Wang, L. Chang, F. Wang, T. Li, and T. Gu, Integrated photonic metasystem for image classifications at telecommunication wavelength, Nat. Commun. **13**, 2131 (2022).

[9] T. Fu, Y. Zang, Y. Huang, Z. Du, H. Huang, C. Hu, M. Chen, S. Yang, and H. Chen, Photonic machine learning with on-chip diffractive optics, Nat. Commun. **14**, 70 (2023).

[10] O. Kulce, D. Mengu, Y. Rivenson, and A. Ozcan, All-optical information-processing capacity of diffractive surfaces, Light Sci. Appl. **10**, 25 (2021).

[11] G. Lu, J. Qiu, T. Liu, D. Zhang, S. Xiao, and T. Yu, Metasurface-based diffractive optical networks with dual-channel complex amplitude modulation, *J. Lightwave Technol.* **42**, 7282 (2024).

[12] H. Kariyawasam, R. Hettiarachchi, Q. Yang, A. Matlock, T. Nambara, H. Kusaka, Y. Kunai, P. T. C. So, E. S. Boyden, and D. N. Wadduwage, QuATON: quantization aware training of optical neurons, Opt. Express **34**, 1 (2026).

[13] J. W. Goodman, *Introduction to Fourier Optics*, 4th ed. (W. H. Freeman and Company, New York, 2017).

[14] Y. Bengio, N. Léonard, and A. Courville, Estimating or propagating gradients through stochastic neurons for conditional computation, arXiv:1308.3432 [cs.LG].

[15] M. Courbariaux, Y. Bengio, and J.-P. David, BinaryConnect: Training deep neural networks with binary weights during propagations, in *Advances in Neural Information Processing Systems 28*, edited by C. Cortes, N. D. Lawrence, D. D. Lee, M. Sugiyama, and R. Garnett (Curran Associates, Inc., Red Hook, NY, 2015), pp. 3123–3131.

[16] I. Hubara, M. Courbariaux, D. Soudry, R. El-Yaniv, and Y. Bengio, Binarized neural networks, in *Advances in Neural Information Processing Systems 29*, edited by D. D. Lee, M. Sugiyama, U. von Luxburg, I. Guyon, and R. Garnett (Curran Associates, Inc., Red Hook, NY, 2016), pp. 4107–4115.

[17] F. D. Neeser and J. L. Massey, Proper complex random processes with applications to information theory, IEEE Trans. Inf. Theory **39**, 1293 (1993).

[18] J. W. Goodman, *Speckle Phenomena in Optics: Theory and Applications*, 2nd ed. (SPIE Press, Bellingham, WA, 2020).

[19] G. E. Healey and R. Kondepudy, Radiometric CCD camera calibration and noise estimation, IEEE Trans. Pattern Anal. Mach. Intell. **16**, 267 (1994).

[20] European Machine Vision Association, EMVA Standard 1288: Standard for Characterization of Image Sensors and Cameras, Release 4.0 Linear, 16 June 2021 (European Machine Vision Association, 2021).

[21] L. Azzari, L. R. Borges, and A. Foi, Modeling and estimation of signal-dependent and correlated noise, in *Denoising of Photographic Images and Video: Fundamentals, Open Challenges and New Trends*, edited by M. Bertalmío (Springer, Cham, 2018), pp. 1–36.

[22] M. Konnik and J. Welsh, High-level numerical simulations of noise in CCD and CMOS photosensors: Review and tutorial, arXiv:1412.4031 [astro-ph.IM].

[23] Z. Liu, W. Hunt, M. Vaughan, C. Hostetler, M. McGill, K. Powell, D. Winker, and Y. Hu, Estimating random errors due to shot noise in backscatter lidar observations, Appl. Opt. **45**, 4437 (2006).

[24] R. Hamerly, L. Bernstein, A. Sludds, M. Soljačić, and D. Englund, Large-scale optical neural networks based on photoelectric multiplication, Phys. Rev. X **9**, 021032 (2019).

[25] M. Fan, S. Jin, Y. Gu, X. Zhao, N. Bai, Q. Wang, and C. Lu, Joint loss function design in diffractive optical neural network classifiers for high power efficiency, Opt. Express **33**, 7307 (2025).